# Contribution au Phénomène de Mouillabilité en présence D'un tensioactif anionique «SDS»et non ionique«$C_{11}E_5$».

**H. Alla$^a$, S. Freifer$^a$ , C. Medjellel$^a$**

*a. Laboratoire de Modélisation et Simulation. Faculté des Sciences, BP 1505 El M'Naouar Bir el Djir Oran, 31000, Algérie.*

**Résumé :**

*L'étude expérimentale des gouttes d'eau à des concentrations différentes de tensioactif non ionique ($C_{11}E_5$) et anionique (SDS) déposées sur des lames horizontales en verre a été faite. La 1$^{ère}$ étape consistait à mesurer la tension de surface, la 2$^{ème}$ nous a permis de déterminer l'influence de ces tensioactifs sur les paramètres physiques du mouillage (l'angle de contact..).La simulation numérique de l'équation de profil de la goutte nous a permis de valider les résultats empiriques trouvés.*

**Abstract:**

*The experimental survey of the water drops to different concentrations from non ionic surfactant ($C_{11}E_5$) and set down anionic (SDS) on glass horizontal surfaces has been made. The 1$^{st}$ step consisted in measuring the tension of surface, the 2nd allowed us to determine the influence of these surfactants on the physical parameters of wetting ( contact angle ...).The numeric simulation of the equation of profile of the drop allowed us to validate the empirical results found.*

**Mots clefs:** mouillage, goutte, tension de surface, tensioactif, angle de contact.

## Introduction

La science du mouillage est un bon exemple du domaine de la recherche fondamentale. Mais on constate parfois que l'écart reste grand entre la pratique et la théorie. Pendant longtemps, la recherche académique s'est focalisée sur des systèmes idéaux : plans solides, liquides simples. Mais il est important de comprendre à quel point les progrès enregistrés dans ce domaine ont pu donnée une assise à toute recherche en offrant des références indispensables .Le souci reste grand en ce qui concerne le mouillage des surfaces solides par des liquides contenants des additifs en particularité lorsqu'ils sont des tensioactifs qui posent des questions difficiles: adsorption aux interfaces conduisant à une modification du mouillage. Aujourd'hui les tensioactifs non ionique et anionique sont utilisés dans tous les domaines industriels de la détergence domestique ou textile, dans les préparations pharmaceutiques et dans l'agriculture .Leur emploi toujours croissant est dû grâce à leurs bonnes propriétés physicochimique, compatibilité avec les autres agents tensioactifs ainsi que leur facilités d'approvisionnement et leur bon rapport coût/efficacité.

Cette étude a pour sujet l'étude de l'influence de tensioactif sur le phénomène de mouillabilité.
Le choix des tensioactifs était pour le $C_{11}E_5$ c'est le tensioactif de formule chimique
$C_{11}H_{23}$-(O-$CH_2$-$CH_2$)$_5$-OH et le SDS c'est le tensioactif qui a pour formule chimique
$C_{12}H_{25}$-(O-$SO_3^-$ - $Na^+$ ).

## 1. Etude expérimentale

### 1.1 Mesure de la tension de surface

La mesure de la tension de surface des gouttes d'eau contenant différentes concentration de tensioactif non ionique ($C_{11}E_5$ ) et du tensioactif anionique (SDS), effectuée à l'aide d'un tensiomètre à lame.
La concentration micellaire critique CMC est ainsi évaluée à 0.6 % massique pour le cas du tensioactif anionique le SDS et 0.03% massique pour le ($C_{11}E_5$), comme le montre les figures (1) ci dessous :

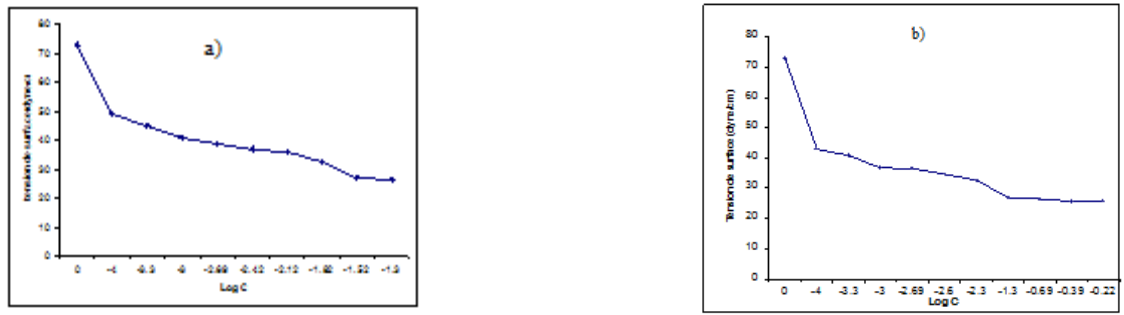

FIG. 1 - Évolution de la tension de surface en fonction de log(c)
pour a) cas du ($C_{11}E_5$) ; b) cas du (SDS).

L'aire de la tête polaire du $C_{11}E_5$ est de 67.46 Å$^2$ et de 289.25 Å$^2$ pour le SDS. De même le calcul de la HLB est de 12.09 dans le cas du tensioactif $C_{11}E_5$ et de 40 pour le SDS. Ceci dit le tensioactif $C_{11}E_5$ est plus hydrophile que lipophile et donc il a une grande affinité dans la phase gazeuse et que le SDS est plus solubilisant.

## 1.2 Mesure des paramètres physiques de la goutte

A une température de 23°C, on a calculé les paramètres physiques des gouttes d'eau (l'angle de contact θ, le rayon η la hauteur H, la masse volumique ρ, la constante capillaire, le nombre de Bond β, la tension de surface σ, et le volume V des gouttes à l'aide d'un appareil de mesure de l'angle de contact G10 (Kruss de hamburg-Germany ) figure (2) ci dessous.

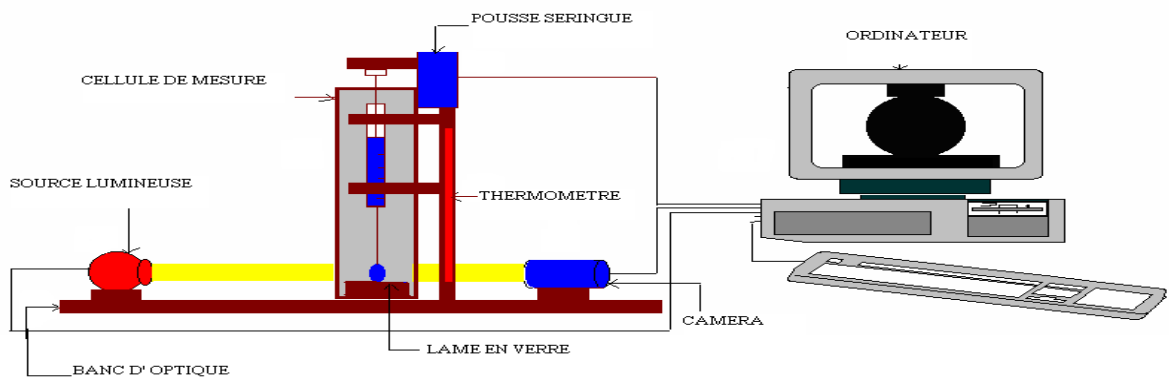

FIG.2 -Schéma descriptif de l'appareil de mesure de l'angle de contact.

Le principe de la mesure de l'angle de contact θ et du rayon consiste à former une goutte de liquide axisymétrique à l'extrémité d'une seringue, dans une surface en verre horizontale, afin que la goutte puisse glisser librement. L'image du profil de la goutte est projetée par un objectif télescopique (caméra CCD 640x480) ; la photo de la goutte ainsi capturée est projetée sur le moniteur, elle ensuite traitée par le logiciel (MOTIC PLUS 2.0), il détermine l'angle de contact, le rayon et la hauteur.

## 1.3 Résultats numérique

Les valeurs expérimentales dans le cas du tensioactif $C_{11}E_5$ sont regroupées au tableau (1) ci-dessous

| Concentration (% massique) | θ (°) | η (cm) | H (cm) | $\rho_1$ (g/cm$^3$) | $\rho_2$ (air) (g/cm$^3$) | σ (dyne/cm) | V (cm$^3$) | Nombre de bond | Constante capillaire (mm$^{-2}$) |
|---|---|---|---|---|---|---|---|---|---|
| 0.05 | 38.7 | 0.1129 | 0.0307 | 1.632 | 0.00119 | 26.4 | 0.0038 | 0.0571 | -0.6059 |
| 0.03 | 41.5 | 0.1086 | 0.0339 | 1.600 | 0.00119 | 27.05 | 0.004 | 0.0666 | -0.4817 |
| 0.015 | 42.5 | 0.1016 | 0.0370 | 1.583 | 0.00119 | 32.21 | 0.0048 | 0.0658 | -0.5798 |
| 0.0075 | 45.9 | 0.1012 | 0.0441 | 1.593 | 0.00119 | 35.70 | 0.0059 | 0.0851 | -0.3153 |
| 0.001 | 49.3 | 0.0945 | 0.0461 | 1.642 | 0.00119 | 40.56 | 0.0067 | 0.0843 | -0.3968 |
| 0.0001 | 52.2 | 0.0923 | 0.0488 | 1.566 | 0.00119 | 48.68 | 0.0083 | 0.0751 | -0.4374 |

TAB.1- Valeurs expérimentales du l'angle de contact θ, le rayon η, la hauteur **H**, la masse volumique ρ, la tension de surface σ, et le volume **V**, nombre de bond dans le cas du **C$_{11}$E$_5$**.

Les valeurs expérimentales dans le cas du tensioactif SDS sont regroupées au tableau (2) ci-dessous

| Concentration (% massique) | θ (°) | η (cm) | H (cm) | $\rho_1$ (g/cm$^3$) | $\rho_2$ (air) (g/cm$^3$) | σ (dyne/cm) | V (cm$^3$) | Nombre de bond | Constante capillaire (mm$^{-2}$) |
|---|---|---|---|---|---|---|---|---|---|
| 0.05 | 39.9 | 0.1177 | 0.0295 | 2.762 | 0.00119 | 27.05 | 0.0021 | 0.0871 | -1.0012 |
| 0.005 | 42.6 | 0.1062 | 0.0304 | 2.773 | 0.00119 | 32.7 | 0.0022 | 0.0783 | -0.8315 |
| 0.0025 | 45 | 0.1035 | 0.0315 | 2.708 | 0.00119 | 34.67 | 0.0024 | 0.0759 | -0.7659 |
| 0.001 | 47.6 | 0.0937 | 0.0402 | 2.788 | 0.00119 | 36.88 | 0.0033 | 0.1199 | -0.7413 |
| 0.0005 | 49.1 | 0.0913 | 0.0417 | 2.750 | 0.00119 | 40.81 | 0.004 | 0.1148 | -0.6607 |
| 0.0001 | 52.4 | 0.0830 | 0.0433 | 2.800 | 0.00119 | 43.03 | 0.005 | 0.1196 | -0.6381 |

TAB.2- Valeurs expérimentales du l'angle de contact θ, le rayon η, la hauteur **H,** la masse volumique ρ, la tension de surface σ, et le volume **V**, nombre de bond dans le cas du **SDS.**

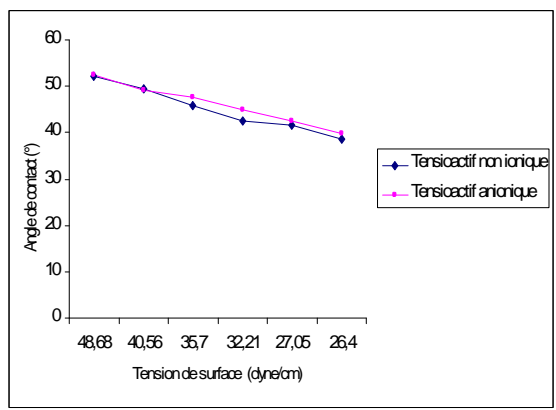 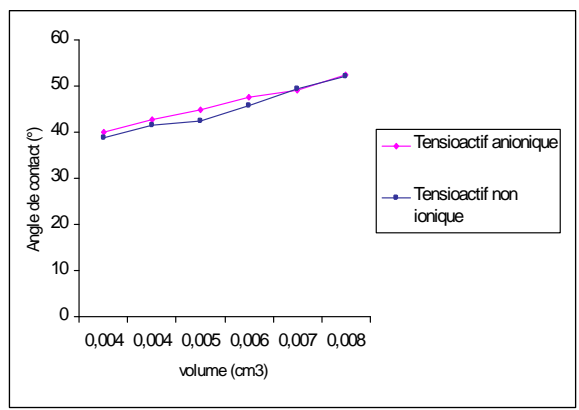

FIG. 3-Evolution de l'angle de contact         FIG. 4-Evolution de l'angle de contact
en fonction de la tension de surface σ.          en fonction du volume V.

La figure (3) présente l'évolution de l'angle de contact des gouttes d'eau contenant de tensioactif en fonction de la tension de surface. On remarque que : l'angle de contact θ décroît avec la tension de surface σ liquide-vapeur pour les cas de tensioactif non ionique et anionique et a partir d'un certain rang Pour une même tension de surface l'angle de contact des gouttes d'eau contenant du SDS est supérieur à celui contenant du $C_{11}E_5$. La figure (4) présente l'évolution de l'angle de contact des gouttes d'eau contenant de tensioactif en fonction du volume de la goutte. On remarque que la courbe est croissante d'où le volume est proportionnel à l'angle de contact en outre, à partir d'une certaine valeur et pour le même volume l'angle de contact du SDS est inférieur à celui du $C_{11}E_5$.

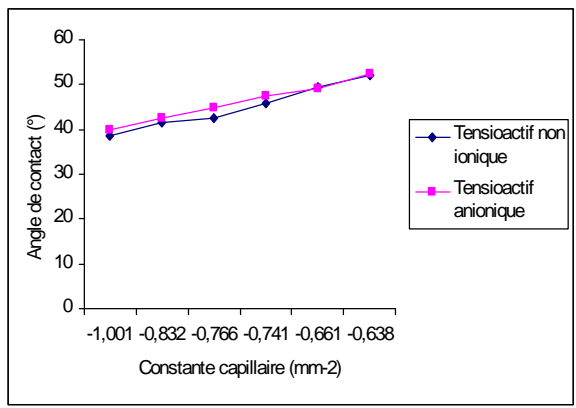
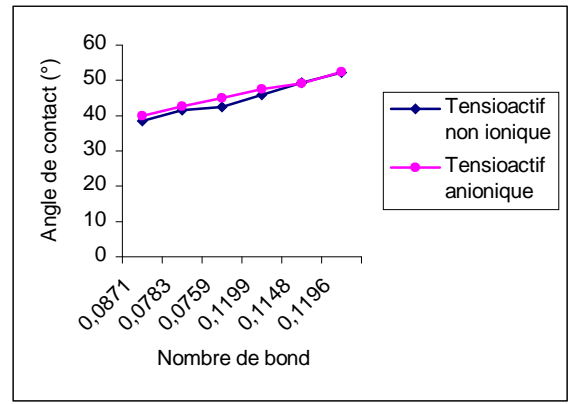

FIG.5-Evolution de l'angle de contact     FIG. 6-Evolution de l'angle de contact
en fonction de la constante capillaire.        En fonction de nombre de bond.

La figure (5) illustre l'évolution de l'angle de contact des gouttes d'eau contenant de tensioactif en fonction de la constante capillaire de la goutte. On remarque que la constante capillaire est proportionnelle à l'angle de contact, et pour une même constante capillaire des deux tensioactifs l'angle de contact du $C_{11}E_5$ est inférieur à celui du SDS. La figure (6) montre bien comment l'angle de contact des gouttes d'eau contenant des tensioactifs varie en fonction du nombre de bond de la goutte. On remarque que le nombre de bond est proportionnel à l'angle de contact, et pour un même nombre de bond en présence des deux tensioactifs l'angle de contact du $C_{11}E_5$ est inférieur à celui du SDS.

## 2. Formulation mathématique

L'équation adimensionnelle (1) du profil de la goutte axisymétrique, posée sur un plan horizontal est une équation différentielle non linéaire de deuxième ordre [1], [4] difficile à exhiber sa solution analytique. Cette équation est tirées de l'équation de Bashforth et Adams [2] et de l'équation de Laplace[3].

$$\frac{y''}{(1+y'^2)^{\frac{3}{2}}} + \frac{y'}{x.(1+y'^2)^{\frac{1}{2}}} - C_1 + C_2.y = 0 \qquad (1)$$

Les conditions aux limites sont : $\begin{cases} y(0) = \dfrac{\eta}{L} \\ y'(0) = -\cot g(\theta) \end{cases}$  (2)

Où, η désigne le rayon de la goutte, L une longueur caractéristique définie par $L = \left(\dfrac{3V}{4\pi}\right)^{\frac{1}{3}}$, où V étant le volume de la goutte, θ étant l'angle de contact au point triple, ainsi $C_1$ et $C_2$ sont données par :

$$\begin{cases} C_1 = L\left(\dfrac{\Delta\rho.g.V}{\pi.\eta^2.\sigma} + 2.\dfrac{\sin\theta}{\eta}\right) \\ C_2 = -\dfrac{\Delta\rho.g.L^2}{\sigma} = -\beta \end{cases} \quad (3)$$

## 2.1 Résultats obtenus

On a regroupé les nouveaux paramètres physiques liés à l'équation (1) dans les tableaux (5) et (6), dans le cas du tensioactif non ionique ($C_{11}E_5$) et anionique (SDS). La simulation numérique de la solution de l'équation (5), munie du système (6) nous a permis d'obtenir les différentes courbes ci-dessous [5], pour chaque cas. (Figure 7)

| Concentration (% massique) | L (cm) | $C_1$ | $C_2$ | $\eta$(cm) | H(cm) |
|---|---|---|---|---|---|
| 0.05 | 0.0968 | 1.6288 | -0.5678 | 1.1663 | 0.3171 |
| 0.03 | 0.0985 | 1.8186 | -0.5626 | 1.1025 | 0.3442 |
| 0.015 | 0.1046 | 2.1369 | -0.5271 | 0.9713 | 0.3537 |
| 0.0075 | 0.1121 | 2.4901 | -0.5497 | 0.9028 | 0.3934 |
| 0.001 | 0.1169 | 2.9836 | -0.5423 | 0.8084 | 0.3944 |
| 0.0001 | 0.1256 | 3.3788 | -0.4975 | 0.7349 | 0.3885 |

TAB. 5- Valeurs expérimentales de la longueur caractéristique L, le rayon $\eta$, $C_1$, $C_2$ *sont* des constantes, et la hauteur H pour des gouttes d'eau contenants un tensioactif non ionique.

| Concentration (% massique) | L(cm) | $C_1$ | $C_2$ | $\eta$(cm) | H(cm) |
|---|---|---|---|---|---|
| 0.05 | 0.0794 | 1.2490 | -0.6312 | 1.4824 | 0.3715 |
| 0.005 | 0.0807 | 1.4454 | -0.5415 | 1.3159 | 0.3766 |
| 0.0025 | 0.0831 | 1.5894 | -0.5314 | 1.2459 | 0.3791 |
| 0.001 | 0.0924 | 2.3683 | -0.6329 | 1.0141 | 0.4351 |
| 0.0005 | 0.0985 | 2.6251 | -0.6411 | 0.9269 | 0.4234 |
| 0.0001 | 0.1061 | 3.5896 | -0.7183 | 0.7823 | 0.4081 |

TAB.6- Valeurs expérimentales de la longueur caractéristique L, le rayon $\eta$, $C_1$, $C_2$ sont des constantes, et la hauteur H pour des gouttes d'eau contenant un tensioactif anionique.

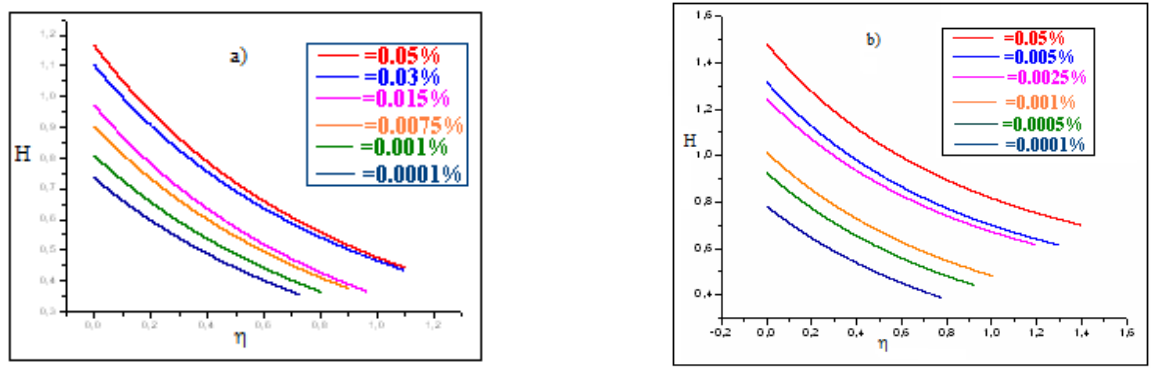

FIG.7-variation de la hauteur H de la goutte en fonction de son rayon η à différentes concentrations
a) cas du ($C_{11}E_5$) ; b) cas du (SDS).

## Conclusions

Ainsi donc, les valeurs trouvées de la concentration micellaire critiques, l'aire de la tête polaire et la HLB pour chaque tensioactifs utilisés (non ionique ($C_{11}E_5$) et anionique (SDS)) confortent bien ceux de la théorie. A une température de 23°C , l'émersion du tensioactif anionique (SDS) l'emporte sur le deuxième tensioactif non ionique ($C_{11}E_5$). La Simulation numérique de la solution de l'équation (1) munie du système (2) donne bien la variation de la hauteur H en fonction du rayon η de la goutte pour des concentrations différentes de tensioactifs variant de 0.0001% à 0.05% , ainsi que l'angle de contact θ correspondant pour chaque cas. Nous avons ainsi donc donner une base de données de paramètres physique liés à la goutte d'eau en présence de tensioactifs non ionique ($C_{11}E_5$) et anionique (SDS) qui pourra intéresser de prêt l'industrie.[6]

## Références